# Development of the poloidal Charge eXchange Recombination Spectroscopy system in Heliotron J

X. X. Lu[1], S. Kobayashi[2], T. Harada[1*, 3], S. Tanohira[1], K. Ida[4], S. Nishimura[4], Y. Narushima[4], D. L. Yu[5], L. Zang[5], K. Nagasaki[2], S. Kado[2], H. Okada[2], T. Minami[2], S. Ohshima[2], S. Yamamoto[2], Y. Yonemura[1], N. Haji[1], S. Watanabe[1], H. Okazaki[1], T. Kanazawa[1], P. Adulsiriswad[1], A. Ishizawa[1], Y. Nakamura[1], S. Konoshima[2], T. Mizuuchi[2]

[1]*Graduate School of Energy Science, Kyoto University, Uji, 611-0011, Japan*
[2]*Institute of Advanced Energy, Kyoto University, Gokasho, Uji, 611-0011, Japan*
[3]*JX Nippon Mining & Metals Corporation, Ibaraki, 319-1535, Japan*
[4]*National Institute for Fusion Science, Toki, Gifu, 509-5292, Japan*
[5]*Southwestern Institute of Physics, Chengdu 610041, China*

**Abstract**

A Charge eXchange Recombination Spectroscopy (CXRS) system designed to measure the poloidal rotation velocity is developed in Heliotron J. The poloidal CXRS system measures the carbon emission line (C VI, n=8-7, 529.05nm) and the Doppler shift of the emission line provides the information of plasma rotation velocity. A high throughput photographic-lens monochromator (F/2.8) with 0.73nm/mm dispersion is adopted to achieve high rotation velocity and temporal resolution. Since two heating neutral beams from two tangential injectors (NBI) are used as the diagnostic beams, a wide observation range ($0.26 \leq r/a \leq 0.92$) is covered by 15 sightlines with a high spatial resolution ($\Delta\langle r/a \rangle < 0.06$) at peripheral region (r/a > 0.6). The system design and the calibration method are presented. The initial results of poloidal rotation measurement show an electron diamagnetic rotation in a NBI heated plasma, while an ion diamagnetic rotation is observed when ECH is additionally applied. The evaluated radial electric field profile shows a positive $E_r$ at plasma core region in the ECH+NBI plasma, and a negative $E_r$ in the NBI heated plasma.

## 1. Introduction

It is widely known that the radial electric field ($E_r$) can affect the heat and particle transport and the MHD activity in magnetically confined plasmas [1, 2], which plays an important role in improved plasma confinement like L-H transition [3, 4]. The radial electric field can be evaluated from the plasma rotation velocity based on the momentum force balance equation [5, 6]. Charge eXchange Recombination Spectroscopy (CXRS) is a powerful diagnostic method to measure ion temperature, impurity density and rotation velocity profiles [7, 8] and has been utilized in many magnetically confined fusion devices such as DIII-D tokamak [9], Joint European Torus (JET) [10], Experimental Advanced Superconducting Tokamak (EAST) [11] and Large Helical Device (LHD) [12]. A parallel CXRS system has been developed to measure the ion temperature and parallel rotation velocity profiles in Heliotron J [13, 14]. In order to evaluate the radial electric field profile [15], it is indispensable to measure the plasma rotation velocity in the poloidal direction.

This article describes a development of the poloidal CXRS system, which designed to measure the radial profile of poloidal rotation velocity with high accuracy. The design study of a poloidal CXRS system is summarized in Chapter 2, the system calibration is described in Chapter 3, the initial results of poloidal rotation velocity measurement and $E_r$ evaluation are reported in Chapter 4 and the summary is in Chapter 5.

## 2. System design of poloidal CXRS for Heliotron J

Figure 1 shows a schematic view of the poloidal CXRS system in Heliotron J, which is a medium-sized (R=1.2m, a=0.2m) helical-axis heliotron device [16]. This system consists of two observation ports, optical fiber sets, calibration lamp, monochromator, CCD camera and data acquisition system. Figure 2 shows a schematic view of the poloidal CXRS system sightlines with magnetic flux surface and vacuum vessel in Heliotron J. A CXRS sightline measures the line integrated emission lights that includes the emission from the NBI beam region and non-NBI region, which is called background region. Here the emission in background region is generated through electron impact excitation and charge exchange with thermal neutral particles [17]. Two observation ports

(to view the beam region and background region, respectively) are installed in toroidally symmetrical positions (toroidal angle = 71.05 deg and 197.05 deg) to remove the emission light from the background region. Here a commercial camera lens (M0814-MP2, C mount) with 8mm focus length and 1.4 F-number is adapted to focus the emission light into an optical fiber bundle installed at the observation port. The optical fiber bundle consists of 15 channels. The numerical aperture (NA) is 0.2 and the core diameter is 0.2mm for each fiber channel. The optical fiber leads the incident light into a photographic-lens monochromator (Bunkoukeiki CLP-400) with high throughput (F/2.8), the focal length of 400mm and a grating of 2160 gloves/mm. The line dispersion of the monochromator is 0.73nm/mm at 529.1nm. A back-illuminated CCD camera (ANDOR DU-897) with Electron Multiplier (EM) is mounted on the monochromator to capture the line emission. The resolution of CCD camera is 512 × 512 pixel, with the area of $16 \mu m^2$ for each pixel, therefor the observation range of wavelength is 526.1-532.1 nm, and the maximum frame speed is 400Hz using a CCD binning method. The exposure of the CCD camera is triggered by an external pulse generator to synchronize with the plasma discharge sequence.

The radial position of the observation region and spatial resolution for each sightline are evaluated based on the CXR emission calculation. The CXR emission intensity is calculated by the following equation [13],

$$I_{CXR} \propto \int n_{imp} n_{beam} \sigma_{CXR} |v| d\omega dl$$

where $n_{imp}$ and $n_{beam}$ are the impurity ion densities and neutral particle density of NBI, $\sigma_{CXR}$ and $|v|$ are the cross section of charge exchange recombination reaction and the relative velocity between impurity ions and neutral particles, $d\omega$ and $dl$ are the solid angle and length of sightline for each view chord, respectively. In Heliotron J, the carbon emission line (C VI, n=8-7, 529.05nm [8]) is measured, since carbon is a major intrinsic impurity in Heliotron J. In this calculation, the density profile of the fully-ionized carbon ion ($C^{6+}$) is assumed to be peaked one using the following formula,

$$n_{C6+}(r/a) = n_{C6+}(0) \times (1 - (r/a)^2)^3$$

since the ionization potential of $C^{6+}$ (490 eV) is close to the typical core electron temperature of Heliotron J. The density profile of neutral beam particles is estimated by a numerical code [13, 18], which calculates the neutral beam trajectory and ionization point by using a Monte-Carlo method. The expected CXR emission distribution for each channel is shown in Fig. 3. The radial position of observation and spatial resolution are evaluated from the averaged value and standard deviation of the emission distribution of each sightline. Figure 4 shows the radial position of observation and spatial resolution for each poloidal CXRS channel. A wide measurement area ($0.26 \leq r/a \leq 0.92$) is covered by 15 CXRS channels at each observation port. The radial resolution is not good ($\Delta \langle r/a \rangle \sim 0.1$) at core region, because of the line integration effect. On the other hand, a good spatial resolution can be obtained in the range of $r/a > 0.6$ where $\Delta \langle r/a \rangle$ is less than 0.06.

## 3. Calibration

To evaluate the ion temperature and rotation velocity with high enough accuracy, absolute wavelength calibration, channel alignment calibration and channel sensitivity calibrations have been carried out.

Line emission from a Samarium ($Sm$) hollow cathode lamp is utilized to determine the absolute wavelength in the CCD image. Figure 5 shows the spectrum lines of Samarium and CXR emission in plasma discharge. Three $Sm$ lines (527.14nm, 528.29nm and 532.06nm) are observed in the CCD image, and the absolute wavelength of the CXR emission is calculated from the position of $Sm$ lines. Two poloidal CXRS channels are connected to the $Sm$ lamp and an in-situ wavelength calibration is performed during the plasma discharge.

To compensate the distortion of the slit image due to grating smile of the system, the slit image is calibrated with the $Sm$ lamp that connected to every CXRS channel. Figure 6 shows the distortion of the slit image as a function of fiber channel number. The maximum shift is about 2.1 pixels, and the distortion is corrected in the rotation velocity calculation.

Two observation ports are used to subtract the emission light from the cold component. Since the CXR emission component is calculated from the intensity difference between the beam region emission and background region emission, the sensitivity calibration between two optical sets are required to obtain the CXR emission spectrum with high accuracy. The sensitivities of the two optical sets are calibrated with an ECH plasma with constant plasma density ($\bar{n}_e \sim 0.6 \times 10^{19} m^{-3}$), since the background emission is only observed at both beam and background region in the ECH plasma. Here the sensitivity ratio between beam and background channel is defined as following equation,

$$I_{Ratio} = I_{Beam}/I_{Background}$$

where $I_{Beam}$ and $I_{Background}$ are the peak intensity of the emission line that measured at beam and background sides, respectively. Figure 7 shows the intensity ratio between the beam and background sides in 15 CXRS channels.

## 4. Initial results of poloidal rotation velocity measurement

In this chapter, we report the initial results of poloidal rotation velocity ($V_\theta$) measurement for a plasma (#68819, t = 270ms) with balanced NBI heating (300+300kW) and an ECH+NBI plasma (#68318, t = 270ms) with same NBI heating power (300+300kW) in Heliotron J. The radial electric field which evaluated from the experimental results of toroidal and poloidal rotation velocities are studied.

Figure 8 shows the radial profiles of electron density ($n_e$), electron temperature ($T_e$) and ion temperature ($T_i$) for these two cases. The line averaged electron density was about $1 \times 10^{19} m^{-3}$ for both cases. Here a peaked density profile is formed in the NBI heated plasma, on the other hand, the density profile is flat in the ECH+NBI plasma. Due to the center-focused ECH, a higher $T_e$ ($\sim 2keV$) is obtained in the ECH+NBI plasma, while $T_e$ for NBI heated plasma is rather low and almost the same value with $T_i$ ($\sim 0.15keV$).

The profiles of $V_\theta$ that measured with the poloidal CXRS system is shown in Fig. 9 (a). The positive direction of $V_\theta$ is defined as the ion diamagnetic direction. For the NBI heated plasma, $V_\theta$ is in the electron diamagnetic direction although its value is small. On the other hand, for the ECH+NBI plasma, $V_\theta$ near the plasma core region (r/a < 0.4) is in the ion diamagnetic direction, and the velocity is small at the peripheral region. Fig. 9 (b) shows the profiles of the parallel rotation velocity ($V_\parallel$) that measured with a CXRS system in which the sightline is parallel to the magnetic field line [14]. Here the co-direction of parallel rotation is in the direction of plasma current which increase the rotation transform. The parallel rotation at the core region ($r/a < 0.4$) is in counter direction whereas a co-direction rotation velocity is observed at peripheral region ($r/a > 0.4$) in the NBI heated plasma. On the other hand, in the ECH+NBI plasma, the parallel rotation velocity at both core and peripheral region moves toward the counter direction compared with that in the NBI heated plasma. Toroidal rotation velocity ($V_\phi$) is required to calculate the radial electric field. Here the toroidal rotation is calculated from the poloidal and parallel rotation data with the following equation:

$$V_\phi = (V_\parallel - V_\theta \sin\alpha)/\cos\alpha$$

where $\alpha$ is the pitch angle and $\tan\alpha = B_\theta/B_\phi$. In this study, the co-direction of $V_\phi$ is defined as the same direction with $V_\parallel$. Since the $B_\phi$ is much larger than $B_\theta$, the pitch angle ($\alpha$) is small and the $V_\phi$ is dominated by the $V_\parallel$. Fig. 9 (c) shows the calculated toroidal rotation velocity ($V_\phi$) profiles in the ECH+NBI plasma and NBI heated plasma. Here the $V_\phi$ profile in both ECH + NBI plasma and NBI heated plasma is almost the same as those of $V_\parallel$, since the $V_\theta \tan\alpha$ component is small.

To study the poloidal rotation, we have evaluated the radial electric field based on the momentum force balance equation with the following equation [6],

$$E_r = \nabla P_i/en_i Z - (V_\theta B_\phi - V_\phi B_\theta)$$

where $\nabla P_i$ is the pressure gradient, Z is the charge number, $V_\theta$ and $V_\phi$ is the poloidal and toroidal rotation velocity, $B_\theta$ and $B_\phi$ is the magnetic field strength in poloidal and toroidal direction, respectively. In this case, the fully ionized carbon ion is treated. And the $E_r$ is determined by the balance of the poloidal rotation component ($-V_\theta B_\phi$), toroidal rotation component ($V_\phi B_\theta$) and pressure gradient component ($\nabla P_i/en_i Z$). Figure 10 shows the $E_r$ with the components of $V_\phi B_\theta$, $-V_\theta B_\phi$ and $\nabla P_i/en_i Z$ in the NBI heated and ECH+NBI plasmas, respectively. Here the value of pressure contributed radial electric field is relatively weak ($|\nabla P_i/en_i Z| < 0.5$ kV/m) at plasma core region since the carbon pressure gradient ($\nabla P_i$) is small. For helical devices like Heliotron J, the toroidal rotation velocity is relatively small due to a large parallel viscosity, compared with the tokamak device. And the radial electric field is dominated by the poloidal rotation velocity since the toroidal rotation contributed radial electric field ($V_\phi B_\theta$) is relatively small than the poloidal rotation contributed radial electric field ($-V_\theta B_\phi$). In the ECH+NBI plasma, a positive $E_r$ with a large shear is observed at the plasma core region ($r/a < 0.35$), corresponding to a large poloidal rotation in the ion diamagnetic direction. On the other hand, the $E_r$ is in negative with a weak shear outside the core region ($r/a > 0.4$). In the NBI heated plasma, the $E_r$ is negative and the shear is rather weak.

Here we have observed a large poloidal rotation velocity in the ion diamagnetic direction and evaluated a positive $E_r$ with large shear at plasma core region in the ECH+NBI plasma. On the other hand, a small poloidal rotation velocity in electron diamagnetic direction with a weak negative $E_r$ is observed in the NBI heated plasma. The poloidal rotation and $E_r$ profile have the same tendency as the previous experiment results in other magnetically confined fusion devices [2, 19], indicating that the poloidal rotation is measured with high accuracy using the new developed poloidal CXRS system in Heliotron J.

## 5. Summary

A poloidal CXRS system designed to measure the radial profile of poloidal rotation velocity with high accuracy has been developed in Heliotron J. The measurement area is $0.26 \leq r/a \leq 0.92$ which covered by 15 channels of sightline and the radial resolution $\Delta\langle r/a \rangle$ is less than 0.06 in the peripheral region ($r/a > 0.6$). The initial poloidal rotation velocity study is performed for an NBI heated plasma and an ECH + NBI plasma. In the ECH + NBI plasma, poloidal rotation in the ion diamagnetic direction is observed at core region, while a weak rotation in the electron diamagnetic direction is observed in the NBI heated plasma. The radial electric field is evaluated from the poloidal and toroidal rotation velocity based on the momentum force balance equation, which shows a positive $E_r$ at the core region in the ECH + NBI plasma, and a negative $E_r$ in the NBI heated plasma.

**Acknowledgement**

The authors would like to thank the Heliotron J technical staff for their support of the experiments. This work was supported by a Grant-in-Aid for Scientific Research from JSPS, Kiban (C) 15K06645 as well as NIFS/NINS under the NIFS Collaborative Research Program (NIFS15KOCH001, NIFS17KUHL079, NIFS12KUHL052), and partly supported by Research Unit for Development of Global Sustainability in Kyoto University as well as Chinese National Fusion Project for ITER under Grant No. 2014GB124003.

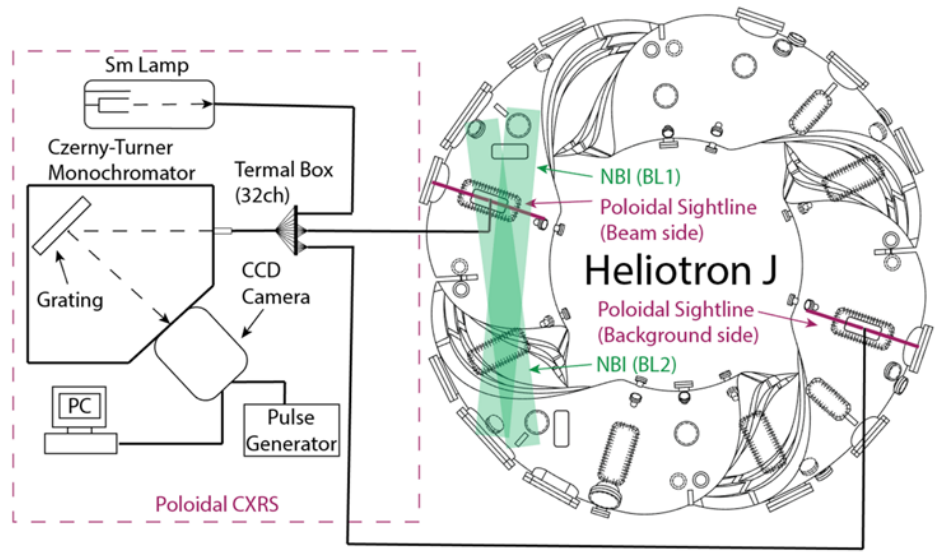

Figure 1. Schematic view of the poloidal CXRS system in Heliotron J.

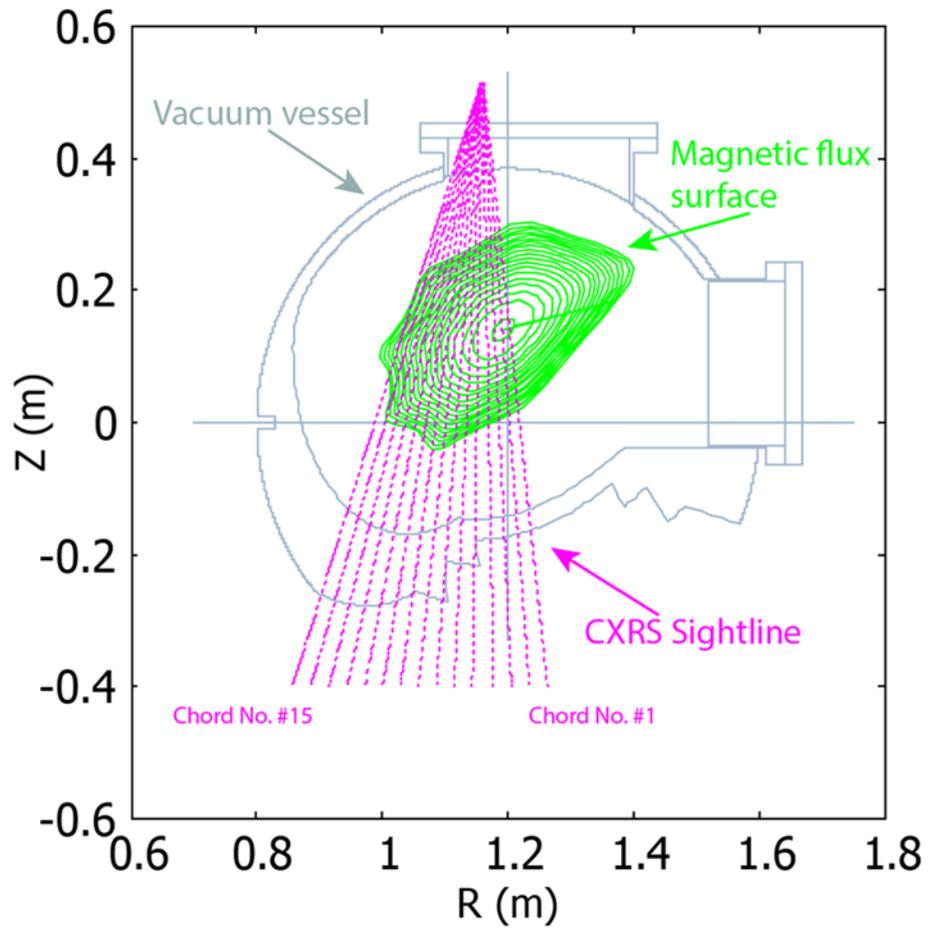

Figure 2. Schematic view of the poloidal CXRS system sight lines, magnetic flux surfaces and the vacuum vessel.

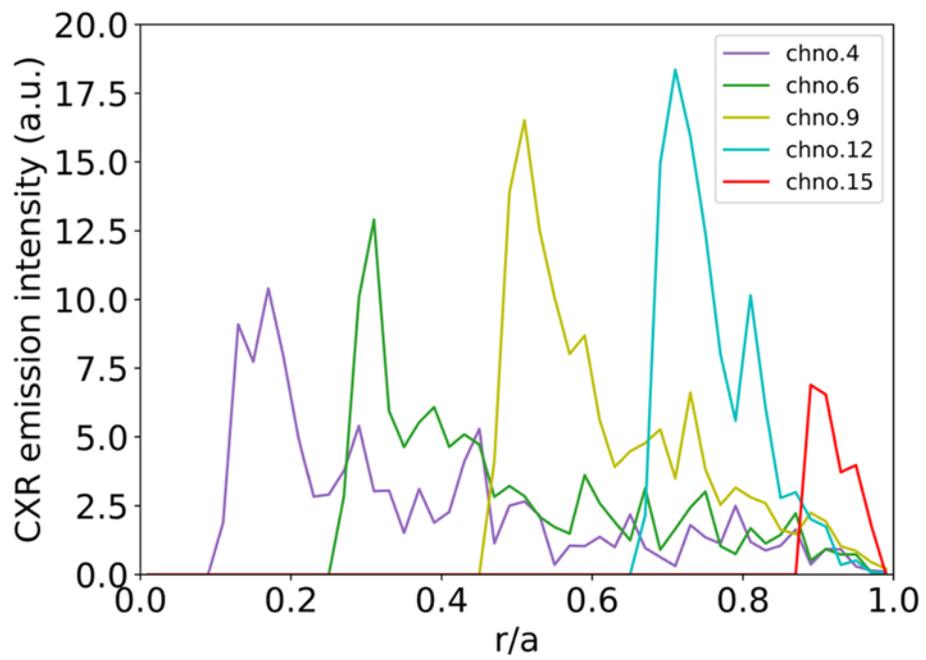

Figure 3. The expected CXR emission distribution of CXR chno. 4, 6, 9, 12, 15 under the standard magnetic configuration

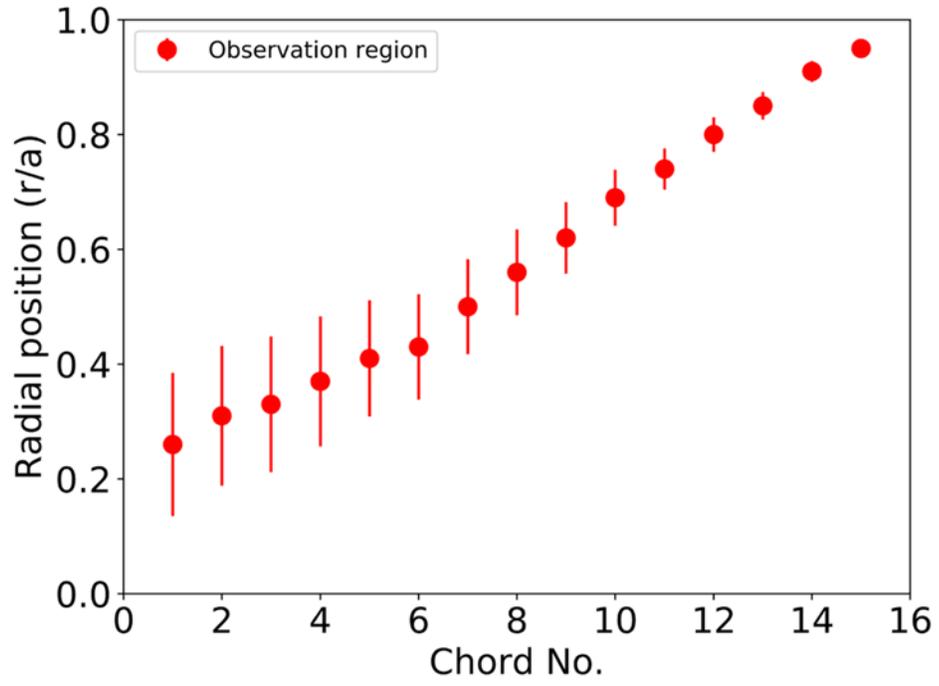

Figure 4. Radial observation point and spatial resolution for each poloidal CXRS sightline

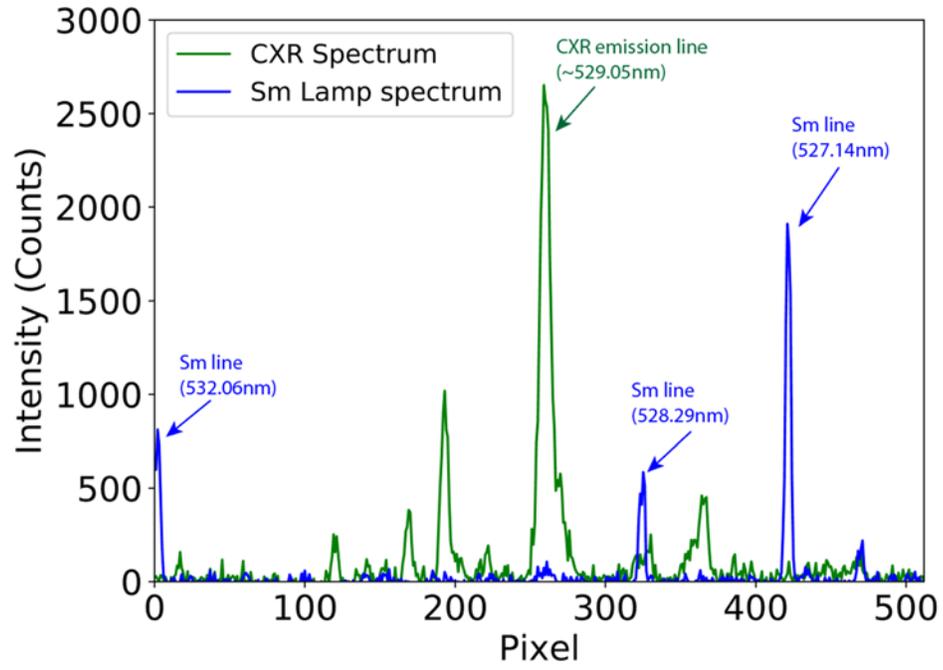

Figure 5. The spectrum line of Samarium and CXR emission in plasma discharge

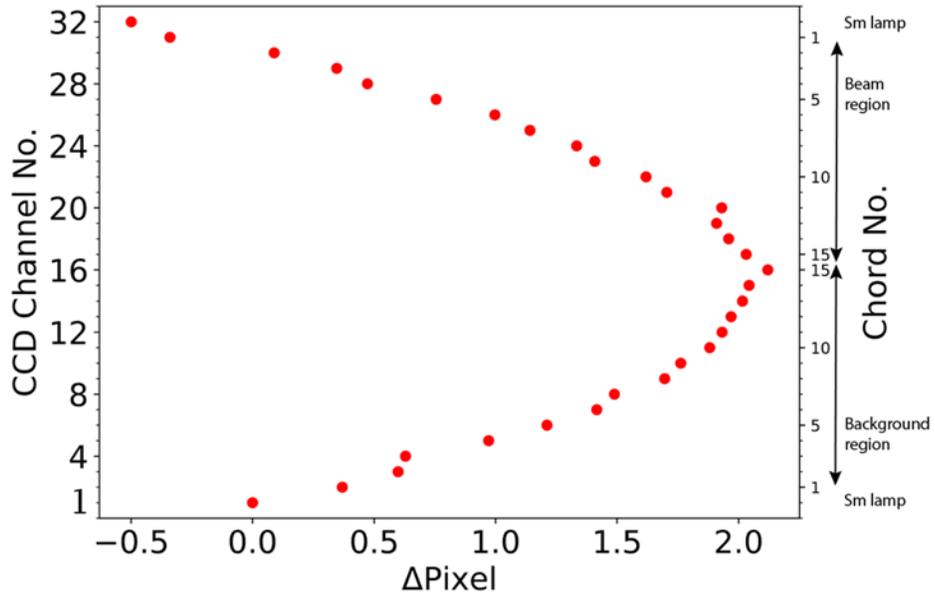

Figure 6. The curve of channel alignment measured with a Samarium lamp

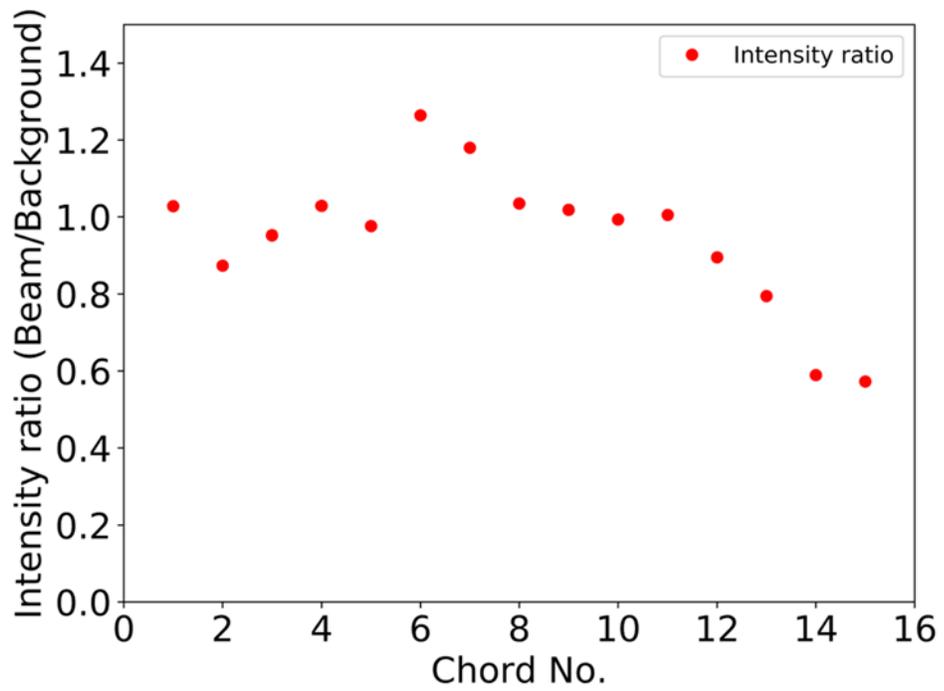

Figure 7. The intensity ratio between beam and background side

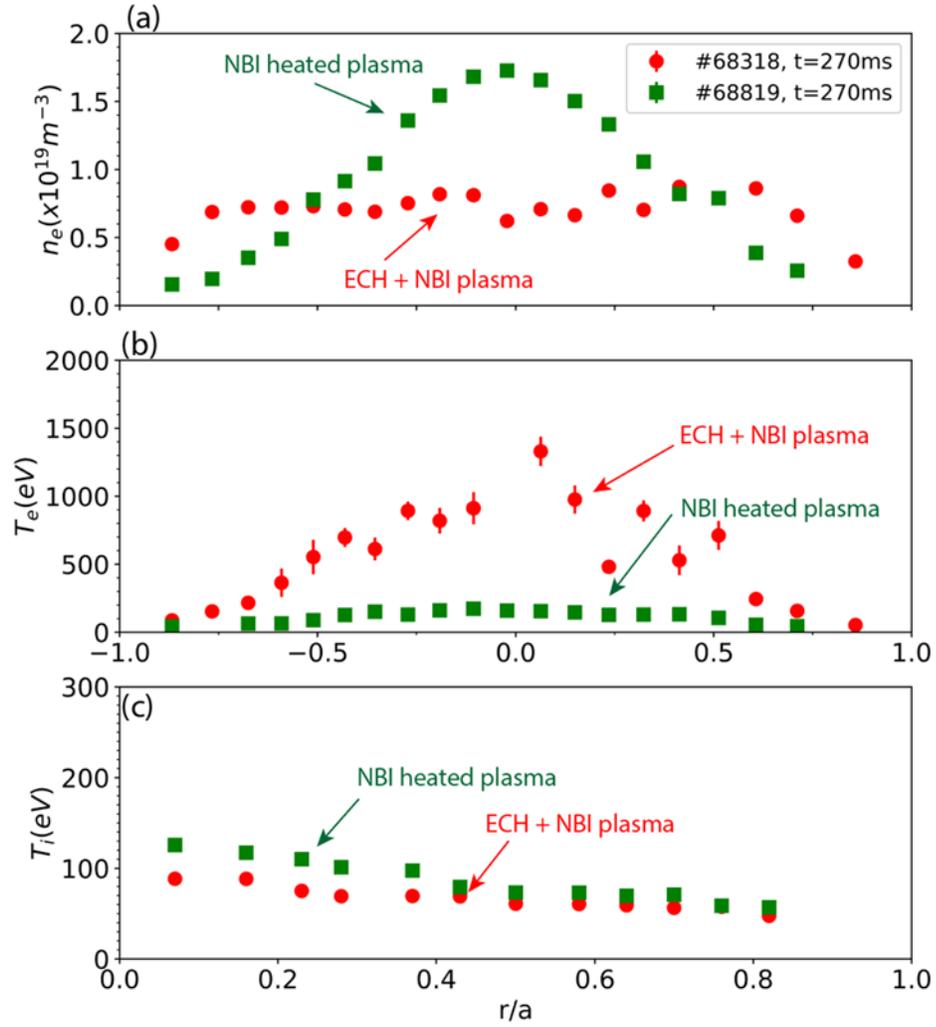

Figure 8. Profiles of (a) $n_e$, (b) $T_e$ and (c) $T_i$ in the ECH + NBI plasma (red) and NBI heated plasma (green)

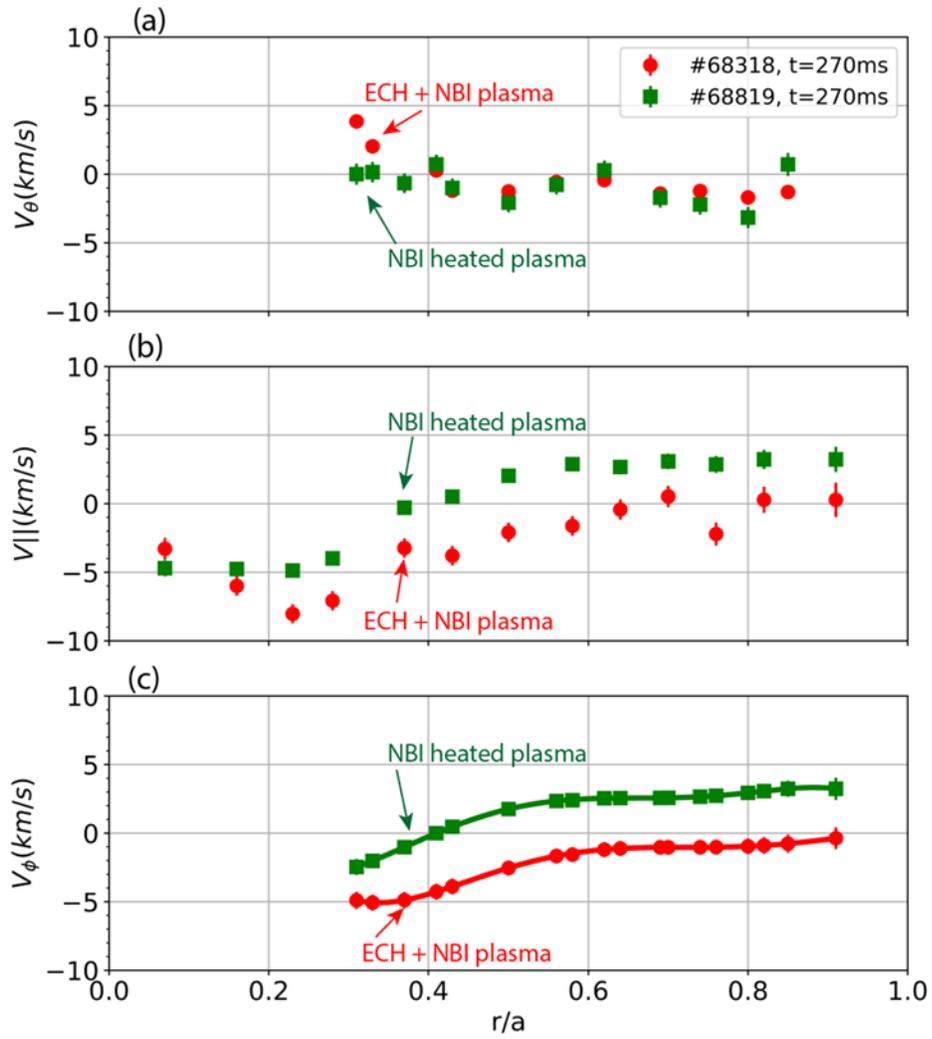

Figure 9. Profiles of (a) $V_\theta$, (b) $V_{||}$ and (c) $V_\phi$ in the ECRH + NBI plasma (red) and NBI heated plasma (green)

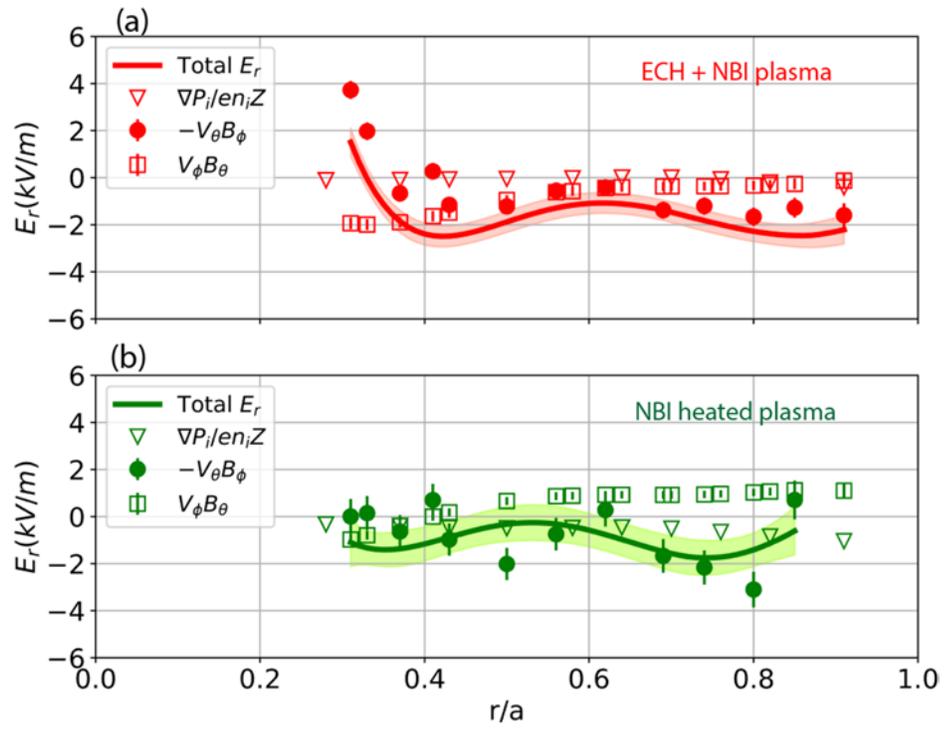

Figure 10. Profiles of $V_\phi B_\theta$ (circle), $-V_\theta B_\phi$ (square), pressure (triangle) contributed $E_r$, and total $E_r$ (solid line) in the case of ECH + NBI Plasma (a, red) and NBI heated plasma (b, green).